# Size-Dependent Charging Energy Determines the Charge Transport in ZnO Quantum Dot Solids

*Morteza Shokrani, Dorothea Scheunemann, Clemens Göhler, Martijn Kemerink[*]*

Institute for Molecular Systems Engineering and Advanced Materials (IMSEAM), Heidelberg University, 69120 Heidelberg, Germany

Keywords: Quantum dots; Charge transport; Hopping; Temperature dependence; Charging energy

**Abstract**

Building up a solid-state material from quantum dots (QD), which are often referred to as artificial atoms, offers the potential to create new materials with unprecedented macroscopic properties. The investigation of the electronic properties of such QD assemblies has attracted attention due to the increasing applications of QD solids in both electronics and optoelectronics. In the past, charge transport in QD assemblies has been explained by a variety of mutually exclusive theories, with the Mott and Efros-Shklovskii variable range hopping models being most common. However, these theories fall short in explaining the anomalous exponents of the temperature-dependent conductivity $\propto \exp(-(T_0/T)^\alpha)$ observed in various QD materials. Here, we measure the temperature-dependent conductivity of semiconducting ZnO QDs under different UV illumination intensity. Regulating the UV intensity allows us to systematically change the effective diameter of the ZnO QDs without having to rely on cumbersome size control by synthesis. Instead, the UV level controls the width of the QD depletion shell and therefore the size distribution in the overall material. We observe exponents that systematically increase from $\alpha = 0.25$ to $\alpha = 0.62$ with increasing illumination intensity, which we interpret in terms of a charge transport being limited by the (size-dependent) charging energy of the QDs.

**Introduction**

Colloidal quantum dot solids are a class of QDs which confine electrons and holes within a particle with diameters below 10nm. The ability to fabricate nanoscale semiconducting or metallic particles along with their unique properties originating from quantum confinement has enabled designing novel materials with properties that are distinct from their bulk counterparts. Therefore, QDs are often referred to as artificial atoms,[1] serving as the building blocks of new materials with engineered functionality. This methodology has proven itself by the growing application of QD materials in diverse fields of research such as (opto-)electronics[2–5] and biology.[6,7]

The macroscopic properties of the QD assemblies can be tuned by carefully selecting the properties of their building blocks such as composition,[8] size[9] and shape[10] of the QDs as well as the type[11] and length[12] of the ligands and packing density. Over the years there has been a lot of research directed toward linking the electronic properties of the QD assemblies with the intrinsic properties of QDs in order to reach the optimized design principles for a specific application.[13] Nevertheless, understanding the significance of each of these parameters poses a great challenge due to possible interplay between them, along with the necessity of maintaining precise control over each of them throughout the chemical synthesis. For instance, there has long been a preference to increase the monodispersity during the fabrication process to minimize the disorder stemming from the size distribution.[14] The perception of size dispersion as a crucial issue in QD solids, along with challenges in systematically controlling the distribution in the synthesis process, has, surprisingly, not resulted in an equally thorough investigation of the effect of the size distribution on charge conduction mechanisms.

Although efforts directed at achieving (narrow) band transport in QD assemblies have been undertaken,[15] these materials are commonly regarded as highly disordered materials with states being localized in real space and energy-space, with charge transport occurring via hopping, that is, thermally assisted tunneling between these states.[16] The charge transport properties of different QD assemblies have been studied previously and most of these investigations reported Mott- or Efros-Shklovskii-like hopping transport mechanisms.[17–19] The temperature dependence of the electrical conductivity in a disordered system can generally be described by:

$$\sigma = \sigma_0 \exp\left(-\left(\frac{T_0}{T}\right)^\alpha\right) \qquad (1)$$

with $\sigma_0$ and $T_0$ being parameters depending on the properties of the film and material (like electronic density of states, Debye frequency of the material, localization length and dielectric constant) and $\alpha$ determining the temperature dependence of the conduction and eventually the type of charge conduction, with $\alpha = 1$ for nearest neighbor hopping (NNH), $\alpha = 1/2$ for Efros-Shklovskii (ES) and $\alpha = 1/(1 + d)$, with $d$ the dimensionality of the system, for Mott-type variable range hopping. However, it is important to keep in mind that these theoretical frameworks of charge transport were established prior to the emergence of QD assemblies, and as such they might be falling short in accounting for novel phenomena associated with them. For instance, there have been reports of anomalous hopping exponents ($\alpha$) between 0.5 and 1 that do not seem to fit into the commonly reported theories.[20–23] Moreover, it is not clear how the found exponents relate to the expected properties of the system in terms of dimensionality or charge density at the Fermi level. Consequently, it is important to acknowledge the necessity for a more conclusive model which could also account for the anomalous exponents in QD assemblies.

An alternative approach to describe charge transport in quantum dot assemblies was followed by Sheng and coworkers.[24] They have developed a theoretical model for the temperature dependence of the charge transport in granular disordered systems and they demonstrate that the whole range of $0.25 < \alpha < 1$ can be explained by the distribution of the Coulomb charging energy of the particles. Interestingly, this line of approach seems to have found little application in the recent experimental literature dealing with charge transport in QDs.

Here we study the temperature dependent electrical conductivity of ZnO QDs assemblies in order to understand the effect of size distribution on the charge transport of QD assemblies. The existence of surface states in ZnO QDs allows us to systematically control the effective diameter (the un-depleted conductive core) distribution by controlling the hydroxyl concentration on the surface of the QDs through illumination with ultra-violet (UV) light. The experimental results cannot be explained by the commonly accepted theories like Mott and ES variable range hopping. We argue that in our assemblies of QDs, and likely in many other QD assemblies with similar size distributions and low or intermediate dielectric constant, the disorder arising from the charging energy distribution plays a crucial role in the energetic landscape of the system. In line with this notion, the experimental data is explained by a charging energy model that attributes the existence of anomalous exponents to the existence of size distribution in the QD assemblies.

**Methods**

Sample fabrication. The ZnO QDs utilized in this study are synthesized by NANOXO employing an organometallic approach and were used as-received. A solution of QDs at a concentration of 2mg/mL in DMSO was drop casted on interdigitated gold electrodes with a distance of 5 μm and dried at 90 ℃ (photos of the sample available in SI).

Electrical measurements. The electrical conductivity measurements were conducted using a 4-point probe method with a Keithley 2636B SMU unit inside a cryostat (current-voltage characteristics available in SI). The temperature sweep was carried out at intervals of 2 Kelvin after sufficiently long time to ensure that the sample has reached the desired temperature. To verify the absence of hysteresis in the temperature dependence of electrical conductivity, measurements were performed during both cooling and heating cycles. To eliminate any potential technical issues arising from poor thermal contact between the substrate and the mounting stage, a thin layer of GE varnish (IMI 7031) was applied to ensure reliability and accuracy of the temperature measurement. The dielectric constant measurements were conducted with an MFIA impedance analyzer from Zurich Instruments. A high-power UV LED (emission wavelength 365 nm) is used to control the effective diameter (hydroxyl concentration) of the QDs. During the temperature sweeps the UV light was turned off to ensure no heating effect arising from the light.

**Results and Discussion**

ZnO is known to be inherently an n-type semiconductor,[25] mostly due to the existence of oxygen vacancies. In bulk ZnO, vacant oxygen sites near the surface are referred to as surface states, which can behave like localized trap states, induce doping and form energetic barriers for interparticle charge transport.[26] Moreover, these surface states exhibit a significant sensitivity to the surrounding atmosphere and consequently exposure to oxygen, moisture and other gases will significantly impact the density of trap states and, concomitantly, the charge transport of the system.[27] In the case of ZnO QDs, this effect gets even more pronounced due to the high surface-to-volume ratio. The surface of the ZnO QDs can chemisorb oxygen species and hydroxyl groups which can trap electrons, which induces a depletion shell inside the QDs, where there are no free electrons to support conduction or to participate in the electrostatic screening that determines the QD's self-capacitive (charging energy) behavior. Consequently, only the inner un-depleted core of the QD remains available for charge conduction. By controlling the concentration of hydroxyl groups on the QD surface, it becomes possible to adjust the diameter of this un-depleted core, which we refer to as the effective diameter.[17] The change in the depletion shell thickness also impacts the (effective) inter QD hopping distance, as a thicker shell increases the separation between conductive cores of adjacent QDs.

The phenomenology of ZnO sketched above has attracted a lot of interest in the fields of chemical sensors, electronics and optoelectronic devices. Therefore, it is commonly believed that the hydroxyl groups can be removed from the surface by exposing the sample to UV light.[28–31] Moreover, the concentration of the hydroxyl groups can be modulated by controlling the UV light intensity, exposure and recovery times, the surrounding atmosphere and temperature.

Figure 1 shows the temporal change of the electrical conductivity of the sample after exposing the film to UV light over more than 24 hours. The electrical conductivity of the thin film increases rapidly within a few minutes and eventually appears to saturate over time. The rise in the conductivity is attributed to the well-documented reduction of hydroxyls and oxygen species on the surface, leading to a decrease in the depletion layer and an increase in carrier concentration and mobility (as discussed previously).[17,26] This process is easily reversible by turning off the UV light. Through careful control of the UV light intensity, temperature, vacuum pressure, exposure and recovery time, we have selected 4 distinct, time-stable concentrations of the surface hydroxyl species, and hence the conductivity, for further investigation of the temperature-dependent conductivity. This enables us to systematically investigate and understand the influence of surface states, and concomitantly the effective inter-dot spacing and dot diameter, on the charge transport mechanism in the QD films without having to perform multiple synthesis and deposition runs.

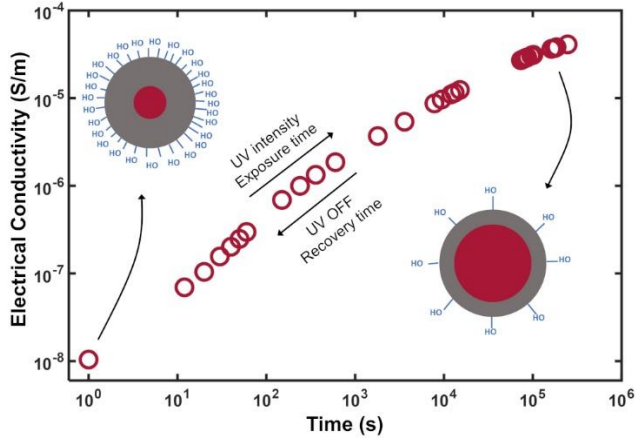

Figure 1. The effect of UV light exposure on the electrical conductivity of a ZnO QD thin film over time. The inset illustrates the reduction of the depletion shell (gray area) as the OH concentration decreases, thereby also increasing the effective QD diameter (red area)

The electrical conductivity of the sample is extracted by 4-point probe current-voltage (IV) measurements at different temperatures (Figure S1). The obtained IV measurements indicate an ohmic and hysteresis-free behavior (Figures S2 and S3). Figure 2(a) illustrates the electrical conductivity of a sample at the 4 different hydroxyl concentrations as a function of temperature. The activated behavior is characteristic of a disordered system. As the OH concentration on the surface increases, the electrical conductivity of the films decreases. This can be explained by the expansion of the depletion shell inside the QDs, hindering the inter-QD charge transport by increasing the inter-dot hopping distance. The initial step in determining the charge transport mechanism is fitting eq. 1 to the experimental data presented in Figure 2 and extracting the exponent. Although this process sounds simple, one has to note that determination of $\alpha$ should be handled with extra caution as some of the conventional methods reported in literature can be insensitive to anomalous exponents that might not follow the typical expected values[32] (Figure S4).

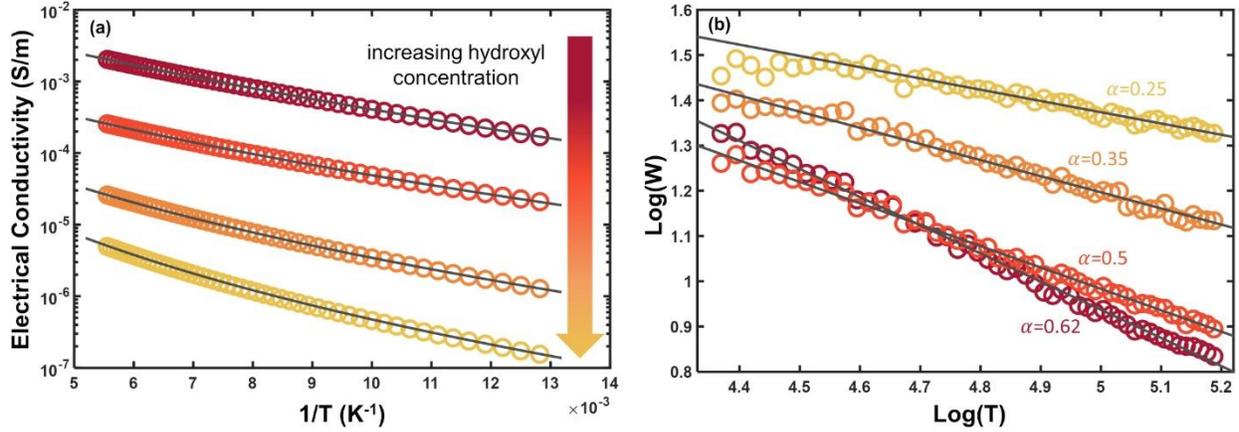

Figure 2. (a) electrical conductivity as a function of 1/T and (b) Zabrodskii plot for different hydroxyl concentrations indicated by the coloring scale. The slope of the fitted lines determines the temperature exponent $\alpha$.

A more reliable approach to determine the optimal exponent involves employing the reduced activation energy, denoted as $W = \frac{d\log\sigma}{d\log T}$. The plotting of $\log(W)$ against $\log(T)$, known as a Zabrodskii plot,[33] is anticipated to yield a straight line for materials exhibiting the stretched exponential behavior versus temperature, with the slope of this line being equal to the temperature exponent in eq 1. Figure 2(b) shows the Zabrodskii plots for the same sample. The slope of the linear fit in the samples from lowest to highest concentration of hydroxyl give 0.624±0.016, 0.473±0.006, 0.375±0.024 and 0.247±0.008, respectively. This indicates that as the depletion shell grows (due to more OH at the surface), the exponents get smaller. The continuously changing exponent, which cannot sensibly be captured by either the ES- or Mott-type expressions that predict fixed ($\alpha = 1/2$) or discrete ($\alpha = 1/(1+d)$) exponents, demonstrates why the widespread analysis method using a linear plot of $\ln\sigma$ against $T^{-\alpha}$ (Figure S4) cannot necessarily distinguish between exponents of 0.5 and 0.62. Complementary to the Zabrodskii-method, the R-square method proves quite reliable for determining $\alpha$ values ranging between 0.25 and 1, with further details available in the Supplementary Information (Figure S5). Following the process of determining the most suitable exponent for fitting the experimental data, we now proceed with identifying the potential reason and model behind the anomalous exponents extracted from the fittings.

One can in principle attribute the ~0.25 and ~0.5 exponents to the commonly used Mott VRH and ES hopping models in 3D, respectively. In this case one would indeed expect the larger exponent, i.e. ES-behavior, to occur at higher charge carrier concentrations, i.e. at longer UV illumination. However, the exponent should then saturate for further illumination, which it does not. The two additional exponents of 0.62 and 0.37 cannot be explained within the aforementioned theories. Each of these 'deviating' exponents might be justified individually by assuming a certain temperature dependence in the pre-exponential parameter of eq 1 ($\sigma_0$) in a way that results in the observed exponents. While a weak temperature dependence in the pre-exponential term has been reported[34], it is generally believed that that the pre-exponential factor has only a weak temperature dependence which can be ignored as compared to the exponential term. In addition, and more importantly, each of the four measurements in Figure 2 would then require unphysical special pleading for a unique temperature dependence of $\sigma_0$ to explain the observed $\alpha$ values.

Another explanation for the anomalous exponents could be the coexistence of, and transition between, two or more different conduction mechanisms. In our case, one could argue that the crossover from ES hopping ($\alpha = 0.5$) to NNH ($\alpha = 1$) potentially yields an exponent of 0.62. Similarly, the crossover between Mott 3D-VRH ($\alpha = 0.25$) and ES hopping could result in an exponent of 0.35. The common characteristics of such crossovers, as reported in literature,[35,18] involve the inability to explain the entire temperature range with a single model (exponent). Typically, the lower temperature range is explained with ES, while the higher temperature range is described with Mott-VRH or NNH. This kind of behavior is then easily visible in Zabrodskii plots of such systems with two distinct regimes. However, in our system the entire temperature range can be adequately fitted with a single exponent, and we were unable to distinguish any potentially two distinctive regimes in our Zabrodskii plots, leading us to exclude any mechanism crossover.

The temperature dependence of the electrical conductivity in assemblies of ZnO nanocrystals with electrochemically gated transistor was investigated by Houtepen et al[21]. Their data was fitted the best by an exponent of 2/3 over a broad temperature range. They proposed "an adaptation of the ES variable-range hopping model by introducing an expression for nonresonant tunneling based on local energy fluctuations, which yields exactly the temperature dependence that is observed experimentally". While the exponent reported in their study is quite similar to the highest exponent observed in our system, the existence of other exponents, especially 0.35, cannot be solely explained by this model. Despite the similarities between the two studied systems (ZnO QDs), it is essential to note that their system is electrochemical, in which for instance the presence of a variable concentration of counter ions adds to the complications in considering the effect of surface states.

Understanding the energetic landscape while accounting for all relevant contributing factors of a potentially disordered system is essential for properly describing the mechanism of charge transfer within that system. In an assembly of QDs with low dielectric constant and diameters below 10 nm, the absolute value of the charging energy of a single QD becomes easily comparable to the other energy scales present in the system. The charging energy $E_c$ in this context is the energy required for adding or removing an electron to a QD and, in first approximation, is equal to:

$$E_c = \frac{e^2}{4\pi\varepsilon_0\varepsilon_r D} \tag{2}$$

where $\varepsilon_0$ is the vacuum primitivity, $\varepsilon_r$ is the relative dielectric constant of the medium and $D$ is the diameter of the QD. For instance, the charging energy in a medium with relative dielectric constant of ~3.5, see Figure S6, and a diameter of 5 nm, vide infra, is equal to 85 meV, that is, significantly above $k_B T$ at room temperature. Taking the charging energy into account it will become apparent that having a distribution in the diameter of the QDs can possibly strongly affect the energetic landscape of the system. More involved expressions for $E_c$ than eq 2 have been proposed, but these would not significantly alter the arguments in the discussion below.[36,37]

Figure 3(a) shows the probability density distribution of the ZnO QDs diameter determined through analysis of atomic force microscopy (AFM) height images of a drop-casted sample from a diluted solution. Although the lateral resolution of AFM is strongly influenced by the diameter and shape of the tip, the vertical resolution (typically sub-nanometer) is independent of that. Since the apparent topographic height of well-isolated QDs, cf. inset to Figure 3a, is not affected by tip convolution effects, AFM provides an effective method for assessing the size distribution of QDs. It is essential to acknowledge that this method is susceptible to errors as the measured QD diameters also include some contribution from the ligands, even though their softness will make this significantly less than double their length, and should lie below ~1 nm. Additionally, the presence of few QDs with diameters above 10 nm can be attributed to the agglomeration of smaller QDs. All these factors can potentially skew the size distribution towards higher diameters. Nevertheless, the majority of our ZnO QDs exhibit diameters below 10 nm.

Figure 3(b) displays the calculated charging energy by eq 2 using the measured diameters and dielectric constant at 78K (see Figure S6). The absolute value of the charging energy spans from very small values of a few meV (from larger QDs) to values around 100 meV (from smaller QDs).

Notably, not only the absolute values but also the energetic disorder (~50 meV) stemming from the charging energy distribution are comparable to, or larger than $k_B T$ and other potential sources of energetic disorder within our temperature range. In this context, it is important that we are in a regime of intermediate quantum confinement, meaning that quantum confinement effects are finite but do not dominate over charging effects, especially not for the larger dots that, as shown below, dominate the charge transport and for which the charging energy distribution (on the order of a few times $k_B T$) remains as the dominant source of disorder. A more detailed discussion of this can be found in the SI.

As we have stated above, in ambient conditions, the surface of the ZnO QDs is populated by hydroxyl groups, which act as traps for charge carriers. Based on the conservation of the charge one can formulate:

$$V_d \times n_e = A_s \times n_{OH} \tag{3}$$

where $V_d$ is the volume of the depleted region, $n_e$ is the electron density inside the QD prior to the adsorption of OH, $A_S$ is the surface area of the QD, and $n_{OH}$ is the hydroxyl concentration on the surface of the QD. Assuming a spherical geometry for QDs, one can write eq 3 as:

$$\tfrac{4}{3}\pi(R^3 - r^3) \times n_e = 4\pi R^2 \times n_{OH} \tag{4}$$

Where $R$ is the total radius of the QD, and $r$ is the radius of the non-depleted core, see inset of Figure 3b. By simplifying eq 4 we get:

$$\frac{R^3 - r^3}{3R^2} = \frac{n_{OH}}{n_e} = \mu \tag{5}$$

Where $\mu$ can be denoted as the parameter governing the width of the depletion shell (S). By increasing the hydroxyl concentration (higher $\mu$), the depletion region expands, thereby reducing the effective diameter of the QDs. Figure 3c shows the effective diameter distribution, calculated from the experimental data in Figure 3a using eq 5 with $\mu = 1$ nm. This calculated size distribution represents a higher OH concentration compared to Figure 3a in which the effective diameter is smaller and therefor the size distribution is shifted towards smaller diameters. The calculated depletion layer thickness is shown in figure S7. Inserting the diameters from Figure 3c in eq 2 results in the charging energy displayed in (Figure 3d). The energetic disorder extracted from the fitted log-normal distribution of the charging energy (solid gray line) shows an increase from 50 to 80 meV, indicating the increasing significance of the charging energy as the OH concentration rises.

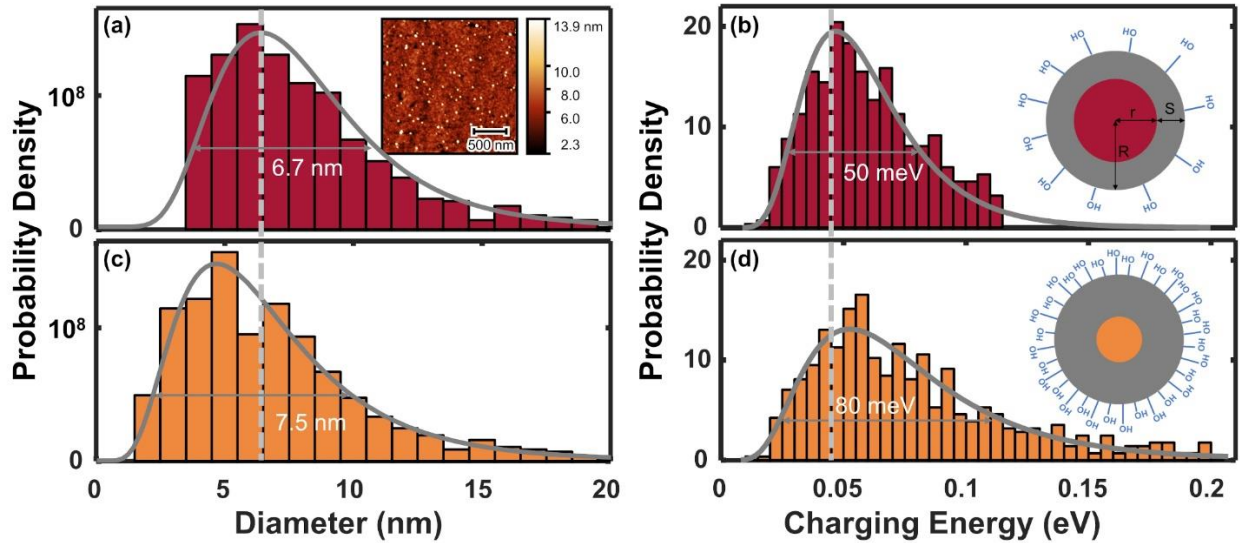

Figure 3. Probability density distribution of diameter and the resulting charging energy of QDs for (a), (b) $\mu = 0$ nm and (c), (d) $\mu = 1$ nm respectively. The inset in (a) shows the AFM image of a diluted sample used for size distribution measurement. The dashed gray line shows the shift in the maximum value of the size and charging energy distributions. The gray solid lines represent log-normal distributions fitted to diameter and charging energy.

To validate that the charging energy distributions in Figure 3 are consistent with the experiments, we calculated the temperature-dependent activation energy for the different OH concentrations by taking by the numerical derivative of $\log \sigma$ to $1/T$. The result is shown in Figure 4. The absolute values are in the same range as Figure 3b, d; the fact that they correspond to the lower half of the shown distributions is expected as the percolating path will preferentially use the easier hops, i.e., those involving lower charging energies. As more thermal energy becomes available, the characteristic or critical hop in the percolating path shifts to larger activation energies, in line with the observations in Figure 4. As predicted by Figure 3b, d, increasing the OH concentration results in higher activation energies as at higher OH concentration the size distribution shifts toward smaller effective diameters and larger charging energies are required to form a percolating path. Figure S8 confirms that the data in Figure 4 cannot be described by the Mott- and ES-VRH models (eq. 1 with $\alpha = 1/4$ and $1/2$, respectively), as expected on basis of Figure 2b.

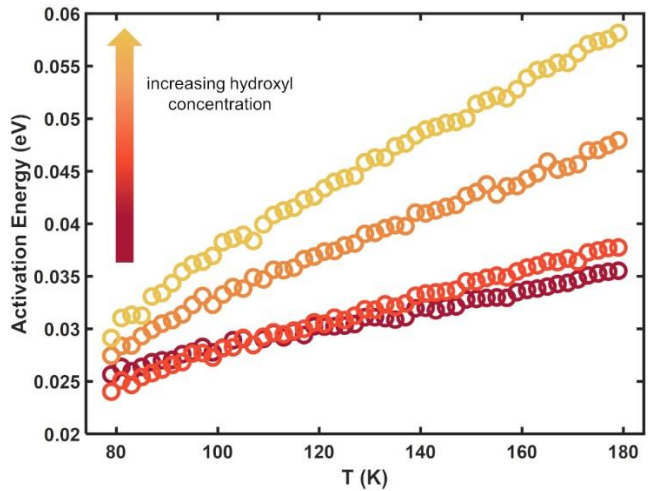

Figure 4. Activation energy versus temperature calculated by the numerical derivation of $\log \sigma$ with respect to $1/T$.

The different exponents in Figure 2b can now be associated with different size and charging energy distributions. For this, we turn to the works of Sheng et al.[24], who conducted a theoretical study on the temperature dependence of the conductivity in disordered granular metals and found that, depending on the size and density distribution of the particles and the temperature range, the temperature exponent of the electrical conductivity can vary between 0.25 and 1. Their findings indicate that, within a constant temperature range, assemblies with smaller grains (higher charging energies) tend to yield smaller exponents while assemblies with bigger average grain sizes are more likely to result in larger exponents. Qualitatively, our findings fully align with their theory as in our case, by increasing the OH concentration, both the average effective grain size and the exponent get smaller. Later, Mostefa et al[38]. performed a detailed analysis of the hopping conduction in granular metals with size distributions and found that depending on the average size and temperature range it is possible to get essentially 'any' exponent between 0.25 and 1: the authors distinguish low- and high-temperature regimes, with exponents increasing (as $1/4 \rightarrow 1/3 \rightarrow 1/2$ and $1/2 \rightarrow 3/5 \rightarrow 1$ for small→medium→large grains, respectively) with dot size, in agreement with our observations.

**Summary**

We have investigated charge transport in assemblies of ZnO QDs, which offers a suitable platform for understanding the effect of the size distribution on the mechanism of charge transport in QD solids with low or intermediate dielectric constant, as it allows to circumvent the complexities associated with chemical synthesis of multiple material batches that would otherwise be needed to vary the (effective) size distribution. Through careful manipulation of the surface OH concentration, we have systematically modulated the depletion shell surrounding the QDs, thus regulating the effective diameter of the QDs. The analysis of the temperature dependent electrical conductivity revealed a systematic decrease in the temperature exponent, ranging from 0.62 to 0.25, with increasing OH concentrations. Our results show that exponents close to 0.25 or 0.5 do not necessarily indicate the dominance of Mott or Efros-Shklovskii variable range hopping. Instead, we attribute this behavior to energetic disorder induced by the QD size distribution, giving rise to a (UV-dependent) distribution in charging energies. Our findings are in alignment with the theoretical models as proposed by Sheng et al. and later Mostefa et al. regarding the effects of size distribution on the charge conduction of granular systems.[24,38]

At the same time, we note that these theories do not explicitly consider any other sources of energetic disorder. In particular, the presence of ionized impurities or dopants would lead to additional energetic disorder and hence affect the density of states at the Fermi level.[39] We therefore believe that the results in this paper highlight the need for further, systematic investigation of the effect of the size distribution in other assemblies of especially semiconductor QDs, in which charging energies are likely to be non-negligible, to further develop our understanding of the relation between the charge conduction and the microscopic properties of the QDs.

## ASSOCIATED CONTENT

**Supporting Information**. Experimental details including figures of the sample, current-voltage and dielectric constant measurements. Extended details and figures for alternative methods of temperature exponent extraction.

## AUTHOR INFORMATION

**Corresponding Author**

Martijn Kemerink- Institute for Molecular Systems Engineering and Advanced Materials (IMSEAM), Im Neuenheimer Feld 225, 69120 Heidelberg. Email: martijn.kemerink@uni-heidelberg.de

**Author Contributions**

M.S. prepared samples and performed the experiments with help of C.G. All authors contributed to data analysis and interpretation. M.K. conceived the idea and coordinated research with help of D.S. M.S. wrote the manuscript with help from all authors.

**Funding Sources**

This work has been funded by the German Research Foundation (Deutsche Forschungsgemeinschaft, DFG) under Germany's Excellence Strategy via the Excellence Cluster 3D Matter Made to Order (EXC-2082/1-390761711). M.K. thanks the Carl Zeiss Foundation for financial support.

**Notes**

The authors declare no competing financial interest.

Supporting Information to

# Size-Dependent Charging Energy Determines the Charge Transport in ZnO Quantum Dot Solids

*Morteza Shokrani, Dorothea Scheunemann, Clemens Göhler, Martijn Kemerink*[*]

Institute for Molecular Systems Engineering and Advanced Materials (IMSEAM), Heidelberg University, 69120 Heidelberg, Germany

## Contents

# Experimental details

Figure S1 shows the current-voltage (IV) characteristics of the resulting film at different temperatures. The measurements were conducted inside a cryostat under high vacuum ($p = 10^{-6}$ mbar). Other than the intensity of the UV light, exposure and the recovery time, the vacuum level can also affect the electrical conductivity of the sample. When the temperature inside the cryostat goes below ~200 K, the pressure of the vacuum chamber drops almost one order of magnitude. In order to have a constant pressure during the measurement, the temperature range is selected not to exceed 180 K.

To ensure stability of the conductivity over time, we implemented a specific measurement protocol as follows: Initially, the sample was cooled to 78 K, and a temperature sweep was conducted in 2 K increments at even temperature values. At each step, we waited approximately 15 minutes to confirm that the sample had reached thermal equilibrium at the desired temperature before taking measurements. This procedure continued until reaching 180 K. Following this, we performed a second temperature sweep by cooling the sample down, measuring at 2 K increments at odd temperature values. Figure S2 shows the conductivity as a function of temperature for both sweeps, with the first sweep data points in red and the second in blue. The overlap of the red and blue points along a single line (without any hysteresis) clearly demonstrates that conductivity remained stable throughout the measurement process, indicating consistent sample behavior over the timescale required for temperature-dependent measurements. Moreover, as further verification, we measured current as a function of time during one complete temperature sweep (figure S3). As highlighted in the zoomed-in insets, each current step corresponding to a 2 K temperature interval remained stable over the given measurement time (~15 minutes), both at low and high temperature ranges.

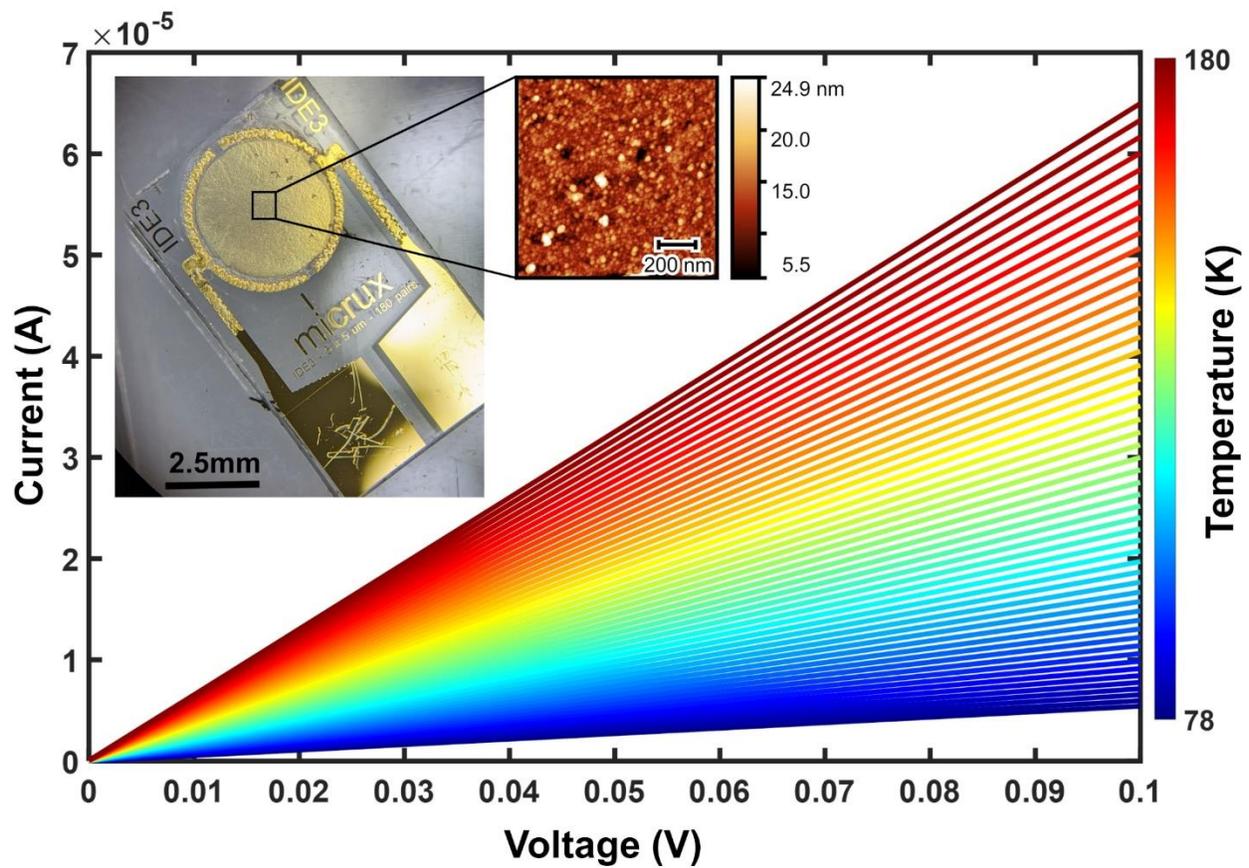

Figure S1. I-V characteristic of the film between 78K to 180K. The inset shows a photo of a sample with an AFM height image from the surface.

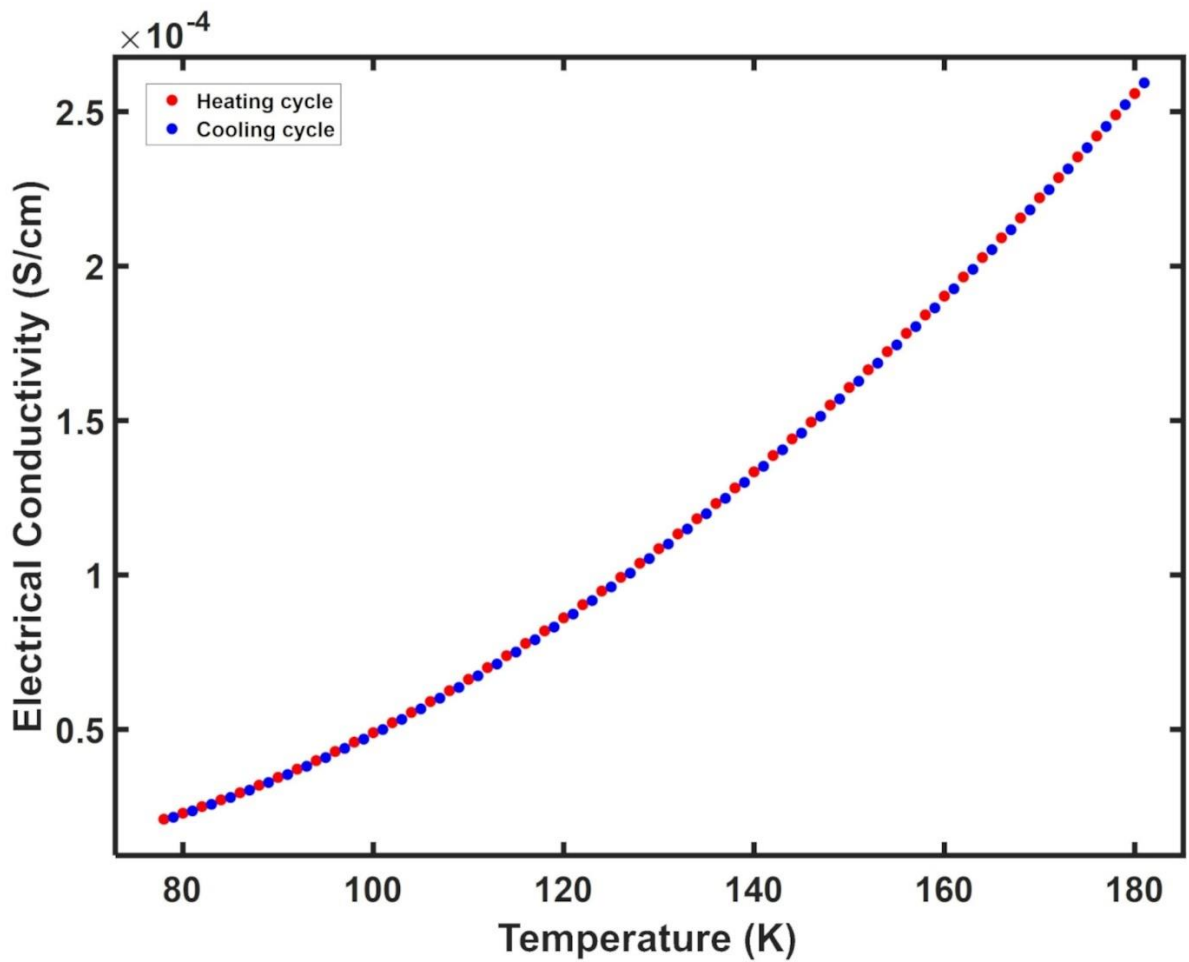

Figure S2. Two temperature sweep cycles between 80 K and 180 K. The red data points represent the heating cycle, measured in 2 K increments at even temperatures, while the blue data points represent the cooling cycle, measured in 2 K increments at odd temperatures. Note the absence of hysteresis, indicating sample stability.

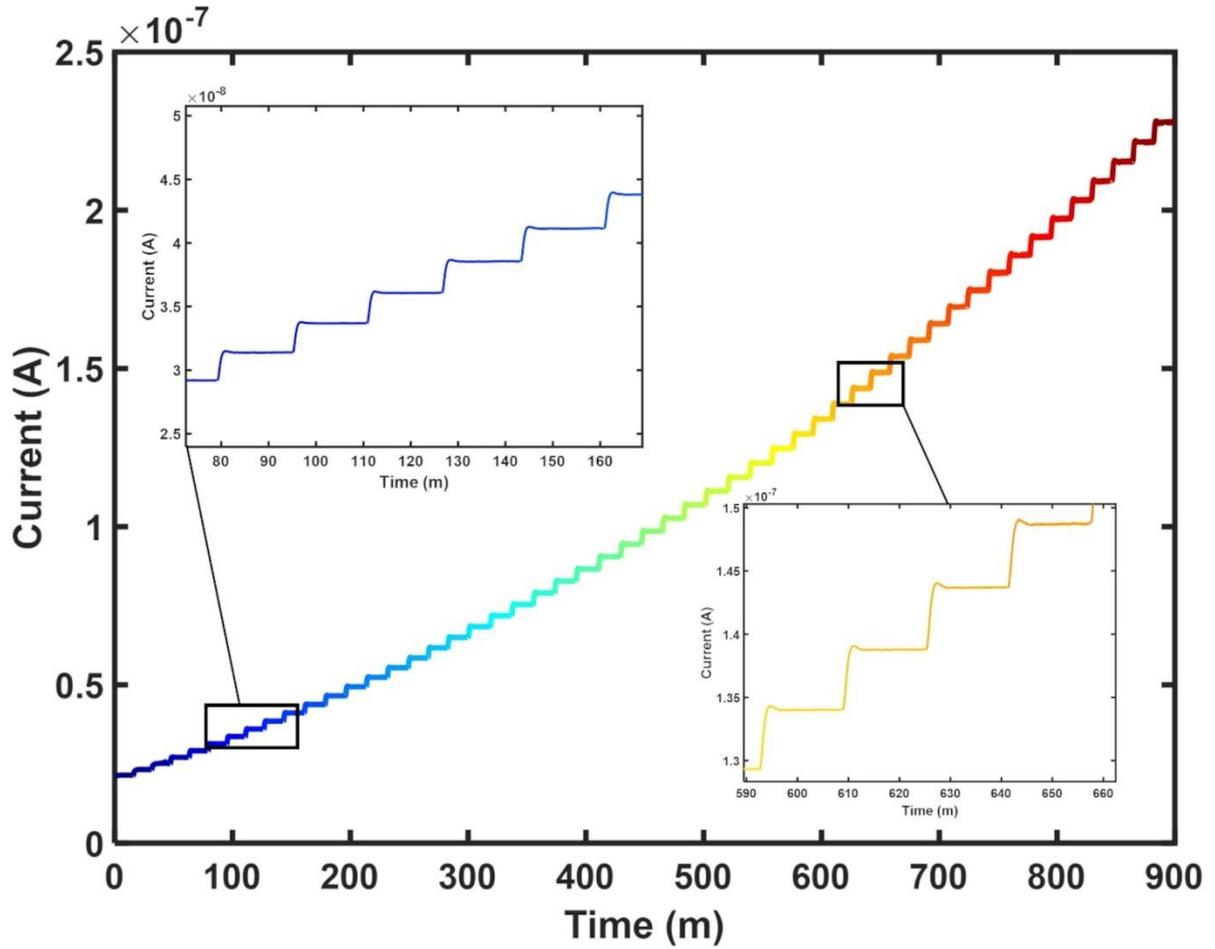

Figure S3. Current versus time for one measurement cycle. Each current step corresponds to a 2 K temperature interval. The inset shows the temporal stability during each measurement.

## Alternative methods to determine the exponent $\alpha$

Figure S4 illustrates the common a practice to select the fitting exponent by visually examining the plotted $\ln \sigma \propto -\left(\frac{T_0}{T}\right)^{\alpha}$ against $T^{-\alpha}$, with $\alpha$ set to 0.25 and 0.5, representing the previously discussed Mott VRH and ES hopping mechanisms, respectively. If the graph displays a visually apparent linearity, it is assumed to indicate that the chosen exponent effectively describes the data. Looking at the fits one can say that the fits are not perfect but still quite reasonable.

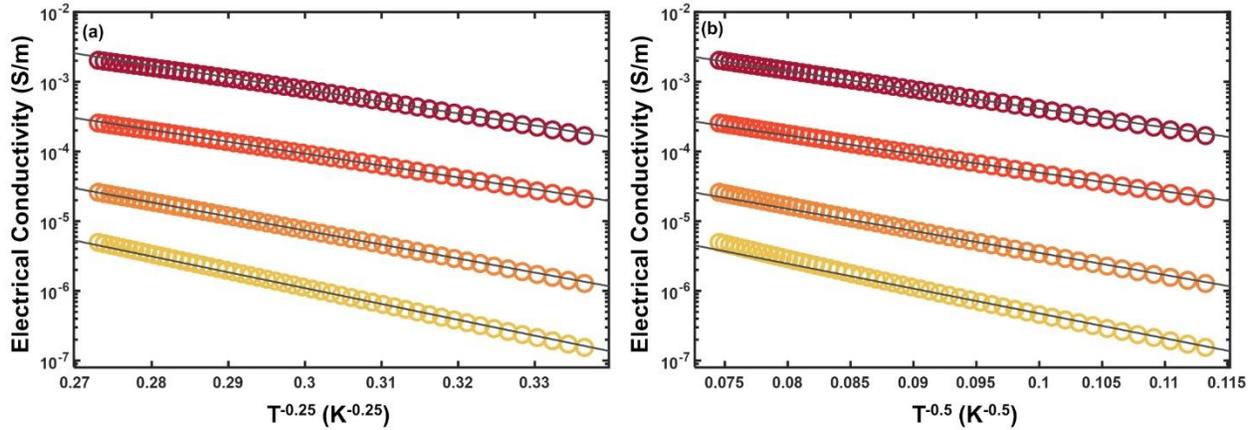

Figure S4. The variation of electrical conductivity as a function of (a) $T^{-0.25}$ and (b) $T^{-0.5}$

Although the reduced activation energy approach gives precise values of the exponent, it becomes hard to fit a linear line in the Zabrodskii plots if the data is noisy. Hence, in addition to that approach, an alternative approach, which involves fitting eq 1 to $\ln \sigma = log\sigma_0 - \left(T_0/T\right)^{\alpha}$ with $\sigma_0$ and $T_0$ as free parameters for a range of fixed $\alpha$ and plotting the R-square of the final fit against $\alpha$, was used. Figure S5 shows the result of this approach for the lowest hydroxyl concentration with the highest R-square achieved at an exponent around $\alpha = 0.6$. The derivative of a parabolic function fitted to the R-square data is plotted as the inset. The point where the derivative is equal to zero is indicated by dashed lines. This point (0.639) corresponds to the maximum of the parabolic fit and hence it represents the optimal exponent which is in good agreement with the exponent of 0.62 extracted from the Zabrodskii plots.

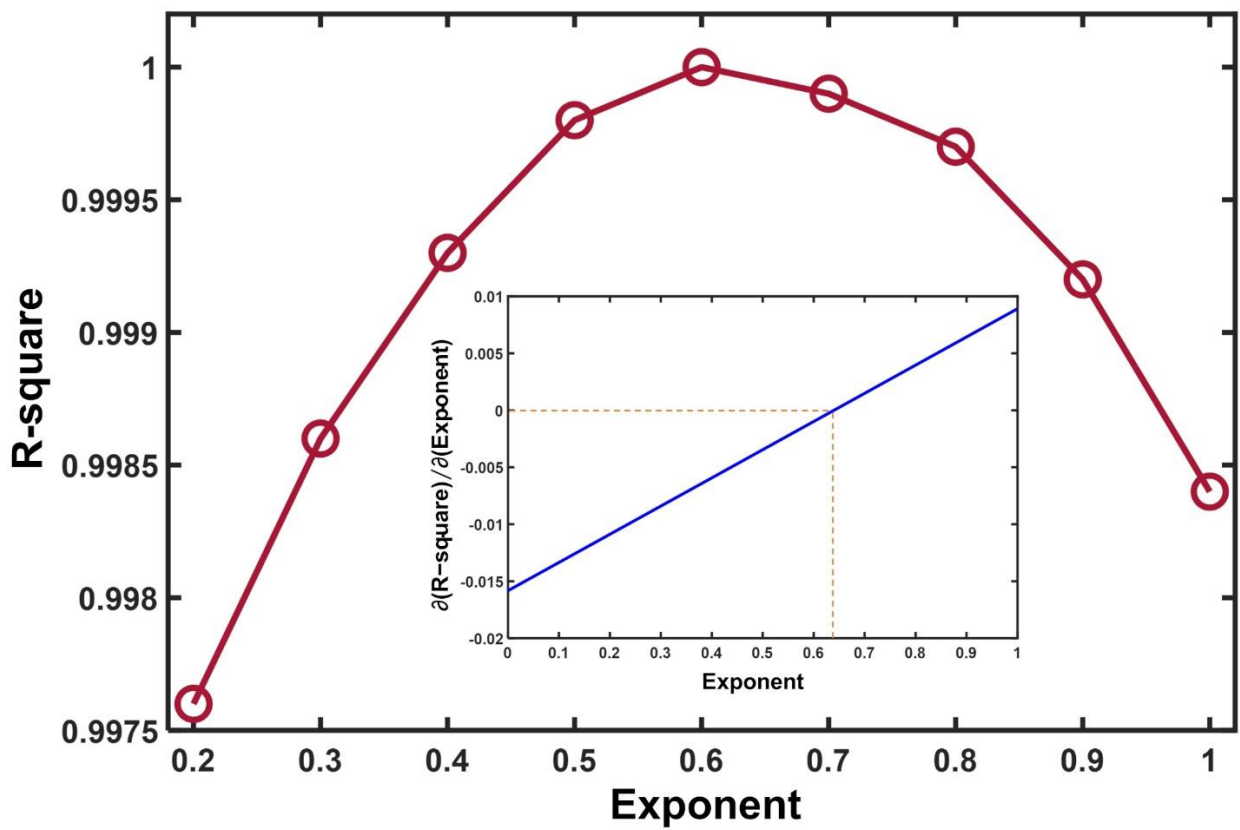

Figure S5. The R-square of the fit to the temperature-dependent conductivity as function of the exponent, as further explained in the SI text above.

# Determining the dielectric constant $\varepsilon_r$

For calculation of the charging energy of the QDs it is needed to measure the dielectric constant of the sample. This is also part of the reason why the solution of the QDs is drop-casted on interdigitated electrodes with known geometrical properties. For dielectric measurement the capacitance of the sample is measured in the same setup of the electrical conductivity measurement using an MFIA from Zurich Instruments. Although the behavior in Figure S6 is relatively complex, for our current purposes the fact that $\varepsilon_r$ is small ($\ll 10$) at all temperatures suffices.

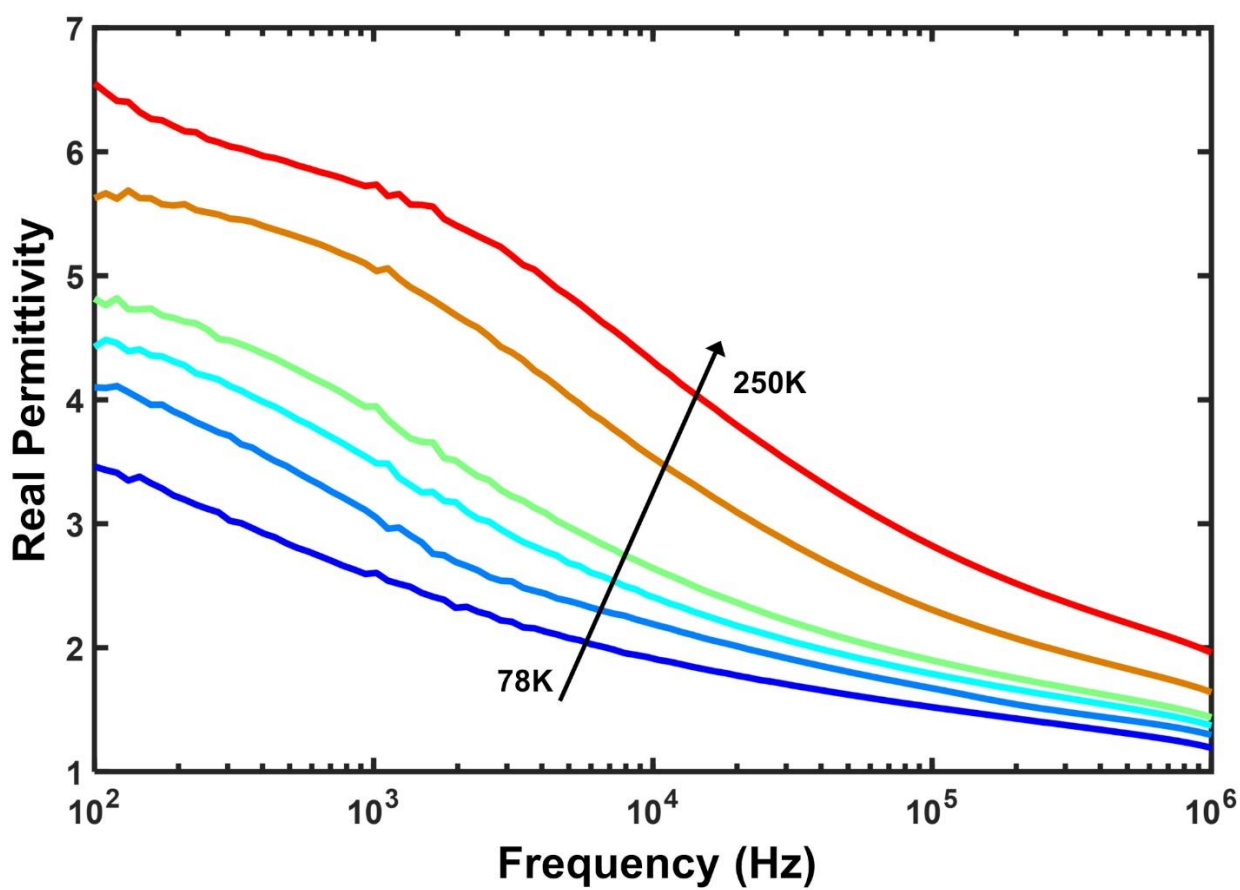

Figure S6. The dielectric constant of the ZnO film with lowest hydroxyl concentration as function of frequency at different temperatures.

# Calculation of the depletion shell thickness

The depletion layer thickness can be calculated based on a charge conservation approach, as mentioned in the main text by equation 3. More specifically, the volume of the depleted region is determined by balancing the charge between the QD's electron density and the hydroxyl concentration on the surface. Assuming a spherical geometry for the QDs, we can define the depletion layer as the difference between the initial QD diameter (figure 3a in the main text) and the effective diameter of the un-depleted core (figure 3c in the main text). The figure below quantifies the depletion thickness for dots with a (geometrical) size distribution as in Fig. 3, using $\mu=1$ nm as in the main text. It is interesting and important to note that the depletion thickness exhibits a distribution (of finite width) due to the variation of the surface-to-volume ratio within this size regime. This means that at a given OH concentration, smaller QDs form relatively thicker depletion layers than larger QDs. In other words, the observed distribution in depletion thickness reflects the initial size distribution of the QDs.

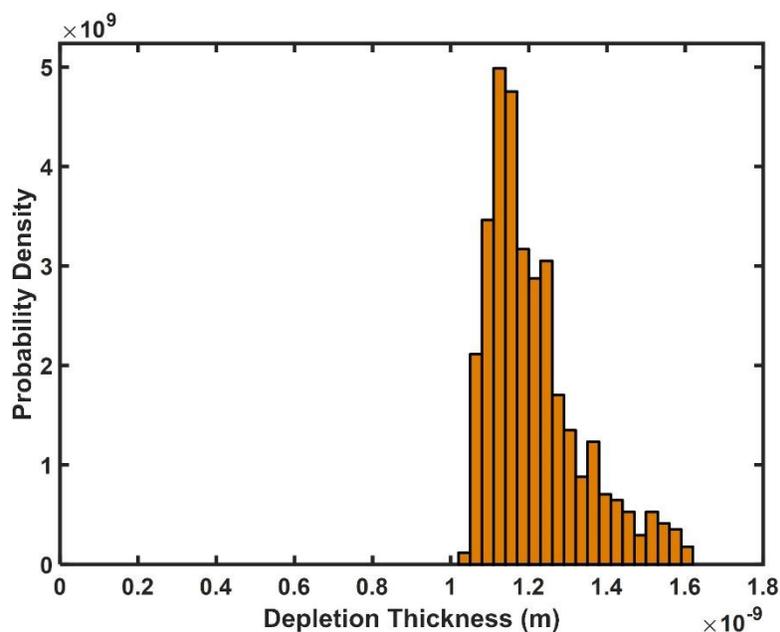

Figure S7. Probability density distribution of depletion shell thickness for $\mu=1$ nm.

# Temperature dependent activation energy

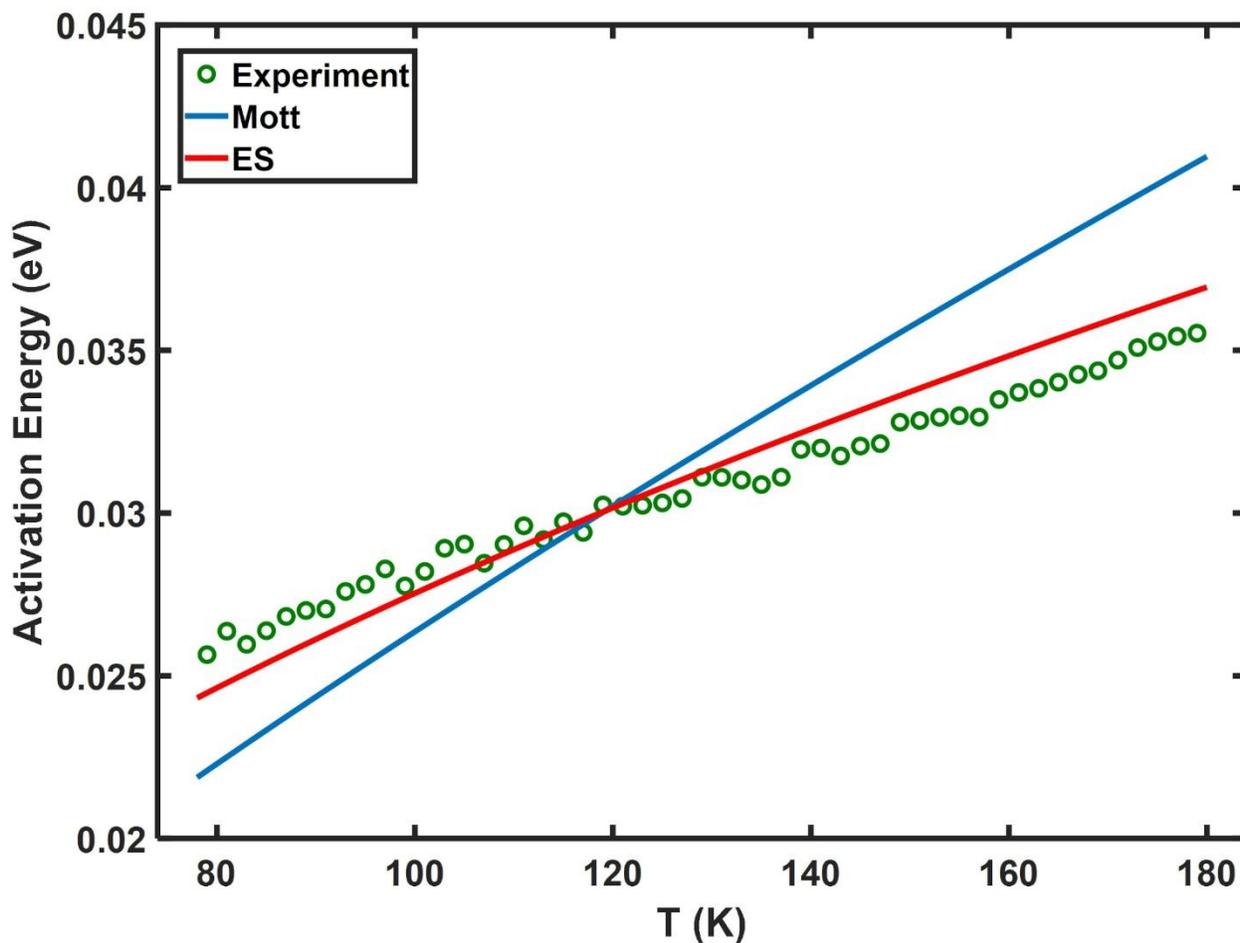

Figure S8. Temperature dependent activation energy of the sample with the lowest hydroxyl concentration (highest conductivity, $\alpha = 0.62$, symbols), compared to the activation energies calculated for the Mott- and ES-VRH fits to the data (eq. 1 with $\alpha = 1/4$ and $1/2$, respectively, lines). As expected from the reduced activation energy plot (Fig. 2), the VRH model fits are inconsistent with the measurements. Similar results are found for the other OH concentrations.

# Exciton Bohr radius in ZnO QDs

An excited electron and a hole in the QD can for a bound pair called exciton. This exciton resembles a hydrogen atom, except that the "nucleus" is a hole instead of a proton. The distance between the electron and hole is referred to as the exciton Bohr radius ($r_B$). the exciton Bohr radius can be expressed in terms of the effective masses of the electron ($m_e$) and the hole ($m_h$), along with the material's dielectric constant ($\epsilon$), reduced Planck's constant ($\hbar$), and the electron charge ($e$):[1]

$$r_B = \frac{\hbar^2 \epsilon}{e^2}\left(\frac{1}{m_e} + \frac{1}{m_h}\right)$$

When the radius of the QD is comparable or smaller than the exciton Bohr radius, the electrons and holes are spatially restricted to the size of the QD and therefore quantum confinement significantly affects the electronic and optical properties of the material. Using the equation above and the material property reported for ZnO,[2] the bulk exciton Bohr radius is expected to be:

$$r_B = \frac{4\pi \times 8.85 \times 10^{-12} \times 8.65 \times \hbar^2}{e^2}\left(\frac{1}{0.24 \times m_0} + \frac{1}{0.59 \times m_0}\right) \cong 2.7 \ nm,$$

which means QDs with diameters smaller than ~5.4 nm experience significant quantum confinement effects, while the influence of quantum confinement diminishes as the particle size increases. Since the mean geometric diameter of our dots is in the same range (cf. Fig. 3) we are likely in an intermediate regime where quantum confinement effects are finite but unlikely to dominate over charging effects that amount to a few times $k_B T$ at room temperature. See also Fig. 4 and its corresponding discussion in the main text.

# Supplementary references